\documentclass[letter]{aa} 

\usepackage[varg]{txfonts}
\usepackage{siunitx}
\usepackage[version=4]{mhchem}
\usepackage{multirow}
\usepackage{graphicx}
\usepackage[caption=false]{subfig}
\usepackage[strict]{changepage}
\usepackage{xcolor}
\usepackage{ulem}
\usepackage[colorlinks=true,citecolor=blue]{hyperref}
\usepackage{float}
\usepackage{amssymb}
\usepackage{amsmath}
\usepackage{lscape}
\usepackage{mathtools}
\usepackage{natbib}
\usepackage{pdflscape}
\usepackage{gensymb}

\defcitealias{zhou2023}{Paper~I}   

\DeclareSIUnit\jansky{Jy}

\newcolumntype{L}[1]{>{\raggedright\let\newline\\\arraybackslash\hspace{0pt}}m{#1}}
\newcolumntype{C}[1]{>{\centering\let\newline\\\arraybackslash\hspace{0pt}}m{#1}}
\newcolumntype{R}[1]{>{\raggedleft\let\newline\\\arraybackslash\hspace{0pt}}m{#1}}

\begin{document}

\title{External tides: an important driver of velocity dispersion in molecular clouds}
\author{J. W. Zhou\inst{\ref{inst1}} 
}
\institute{
Max-Planck-Institut f\"{u}r Radioastronomie, Auf dem H\"{u}gel 69, 53121 Bonn, Germany \label{inst1} \\
\email{jwzhou@mpifr-bonn.mpg.de}
}

\date{Accepted XXX. Received YYY; in original form ZZZ}

\abstract
{Using the 3D density distribution derived from the 3D dust map of the solar neighborhood, the gravitational potential is obtained by solving the Poisson equation, from which the tidal tensor is computed. In the optimal decomposition, the external tidal tensor follows the same formalism as that of a point mass. The average tidal strength of the clouds, derived from both tidal tensor analysis and pixel-by-pixel computation, shows consistent results.  
The equivalent velocity dispersion of the clouds, estimated from the average tidal strength, is comparable in magnitude to the velocity dispersion measured from CO (1-0) line emission. This suggests that tidal effects from surrounding material may play a significant role in driving velocity dispersion within the clouds. Future studies should carefully consider these tidal effects in star-forming regions.}


\titlerunning{}
\authorrunning{}

\maketitle 

\section{Introduction}

One prevailing paradigm about star formation is that the collapse of gas
is driven by gravity, balanced by processes such as turbulence and
magnetic fields \citep{McKee2007-45}. Based on this picture, the virial parameter \citep{Bertoldi1992-395} is always used to determine whether a structure is in a state of gravitational collapse. However, as pointed out in \citet{Ramirez2022-515,Li2024-528}, these studies focus on the role of gravity in individual, isolated parts, neglecting long-range gravitational interactions. Diverse manifestations of gravitational effects on gas within molecular clouds were unveiled in \citet{Li2024-528}. Gravity influences cloud evolution in various ways. Within dense regions, it facilitates fragmentation and collapse. Outside these regions, it can suppress the low-density gas from collapsing through extensive tidal forces and drive accretion. 

Star-forming regions typically exhibit hierarchical gas structures. 
High-mass stars preferentially form in the density enhanced hubs of hub-filament systems \citep{Myers2009,Schneider2012,Motte2018-56,Kumar2020-642,Zhou2022-514}. The hub can drive longitudinal gas inflows along filaments providing further mass accretion \citep{Peretto2013,Henshaw2014,Zhang2015,Liu2016,Yuan2018,Lu2018,Issac2019,Dewangan2020,Liu2021-646,Liu2022-511,Kumar2020-642,Zhou2022-514,Liu2023-522,Xu2023arXiv,Zhou2023-676,Yang2023-953,Zhou2024-686-146}. 
\citet{Zhou2022-514} examined the physical properties and evolution of hub-filament systems in roughly 140 clumps using spectral line data from the ATOMS (ALMA Three-millimeter Observations of Massive Star-forming regions) survey \citep{Liu2020}. We proposed that hub-filament structures, characterized by self-similarity and filamentary accretion, are consistent across different scales in high-mass star-forming regions, spanning from a few thousand astronomical units to several parsecs. This picture of hierarchical, multi-scale hub-filament structures was expanded from the clump-core scale to the cloud-clump scale by \citet{Zhou2023-676}, and later extended to the galaxy-cloud scale in \citet{Zhou2024PASA,Zhou2024-534,Zhou2025-537b}. Previous works have also demonstrated hierarchical collapse and the feeding of central regions by hub-filament structures, as discussed by \citet{Motte2018-56}, \citet{Vazquez2019-490}, \citet{Kumar2020-642}, and references therein.

Gravity is a long-range force. A local gas structure evolves under its self-gravity, but as a gravitational center, its influence can also affect neighboring structures. At the same time, it also experiences the external gravity from neighboring material. The tidal and gravitational fields are mutually interdependent. 
Tidal forces have been proposed in previous studies as a factor that can either regulate or initiate star formation. \citet{Ballesteros2009-393,Ballesteros2009-395} examined the effects of tidal forces induced by the Galactic potential on molecular clouds, demonstrating that these forces can either compress or disrupt the clouds, thereby impacting star formation efficiency. For M 51 and NGC 4429, the models of \citet{Meidt2018-854,Liul2021-505} investigated the influence of the host galaxy potential on molecular clouds, showing that cloud-scale gas dynamics result from the interplay between the galactic potential and gas self-gravity, which plays a key role in shaping molecular cloud properties.  
However, for the Large Magellanic Cloud (LMC), \citet{Thilliez2014-31} found that tidal instability does not hinder star formation. Furthermore, studies by \citet{Dib2012-758,Thilliez2014-31,Zhou2025-537a} indicate that, for molecular clouds in the Milky Way, the LMC, and NGC 628, the shear derived from the galactic rotation curve is negligible. \citet{Ramirez2022-515} also suggested that tidal stresses from nearby molecular cloud complexes contribute more to interstellar turbulence than the overall galactic potential. 
The hierarchical/multi-scale hub-filament structures means the extensive tidal interactions in the interstellar medium (ISM). 
Regardless of whether the galactic potential influences molecular clouds and their complexes, these clouds are inevitably impacted by the cumulative tidal interactions with surrounding material. Such interactions may hinder gravitational collapse, suppress instability growth, and inhibit star formation within the cloud.
\citet{Zhou2025-537a} carried out an examination of large-scale galactic effects on molecular cloud properties in NGC 628, and found the significant impact of tidal effects from neighboring material on the evolution of molecular clouds. 
The tidal effects from neighboring material may also be a significant contributing factor to the slowing down of a pure free-fall gravitational collapse for gas structures on galaxy-cloud scales revealed by velocity gradient measurements \citep{Zhou2024PASA,Zhou2024-534}.
In \citet{Zhou2024-686-146}, the deformation due to the external tides can also effectively slow down the pure free-fall gravitational collapse of gas structures on clump-core scales. These mechanisms could be called "tide-regulated gravitational collapse".

Due to the diffuse and complex morphology of ISM, a complete tidal calculation would be complex. One should derive the gravitational potential distribution from the 3D density distribution and then calculate the tidal field according to the
gravitational potential, as presented in \citet{Li2024-528}. In observations, we can usually only obtain the 2D projected surface density map. 
However, building on the accurate distances enabled in the {\it Gaia} era, we have been able to infer the 3D density distribution of the ISM by the 3D dust
mapping technique \citep{Rezaei2018-618,Chen2019-483,Green2019-887,Lallement2019-625,Hottier2020-641,Leike2020-639,Edenhofer2024-685}. 
In this work, we first calculated and decomposed the tidal field using the 3D dust map, then explored methods to estimate the tidal strength within the clouds induced by external gravity. Finally, we assessed the impact of external tides on the physical properties of molecular clouds.

\section{Data}
\subsection{3D dust map}
\citet{Edenhofer2024-685} offers the most accurate 3D dust map to date for the solar neighborhood within 1.25 kpc, employing a novel Gaussian process prior approach to reduce the impact of fingers-of-god artifacts.
This 3D map was generated using stellar distance and extinction estimates from \citet{Zhang2023-524} derived from {\it Gaia} spectra.
There are 12 samples from their inferred 3D dust extinction distribution. In this work, our calculation is based on the posterior mean of their reconstruction, which is the average of these 12 samples. 
Using the method described in \citet{O'Neill2024-973}, we converted the differential extinction in the 3D map to the number density of hydrogen nuclei, $n$, independent of phase ($n$ = $n_{\rm H_{I}}$+2$n_{\rm H_{2}}$). Then the volume density is, $\rho=1.37  m_{\rm p}  n$, where $m_{\rm p}$ represents the mass of a proton, and the factor 1.37 accounts for the contribution of helium to the total mass, based on cosmic abundance ratios relative to hydrogen.

\subsection{CO (1-0) line emission}
We also utilized the full position-position-velocity ($l$-$b$-$v$) cube of CO \footnote{\url{https://lweb.cfa.harvard.edu/rtdc/CO/}} from \citet{Dame2001-547}. The spatial range of this cube is all Galactic longitudes and $\pm$ 30$^{0}$ Galactic latitude. The velocity range is $\pm$ 320 km s$^{-1}$.

\section{Results}

\subsection{Structural identification}\label{identification}

\begin{figure*}
\centering
\includegraphics[width=0.95\textwidth]{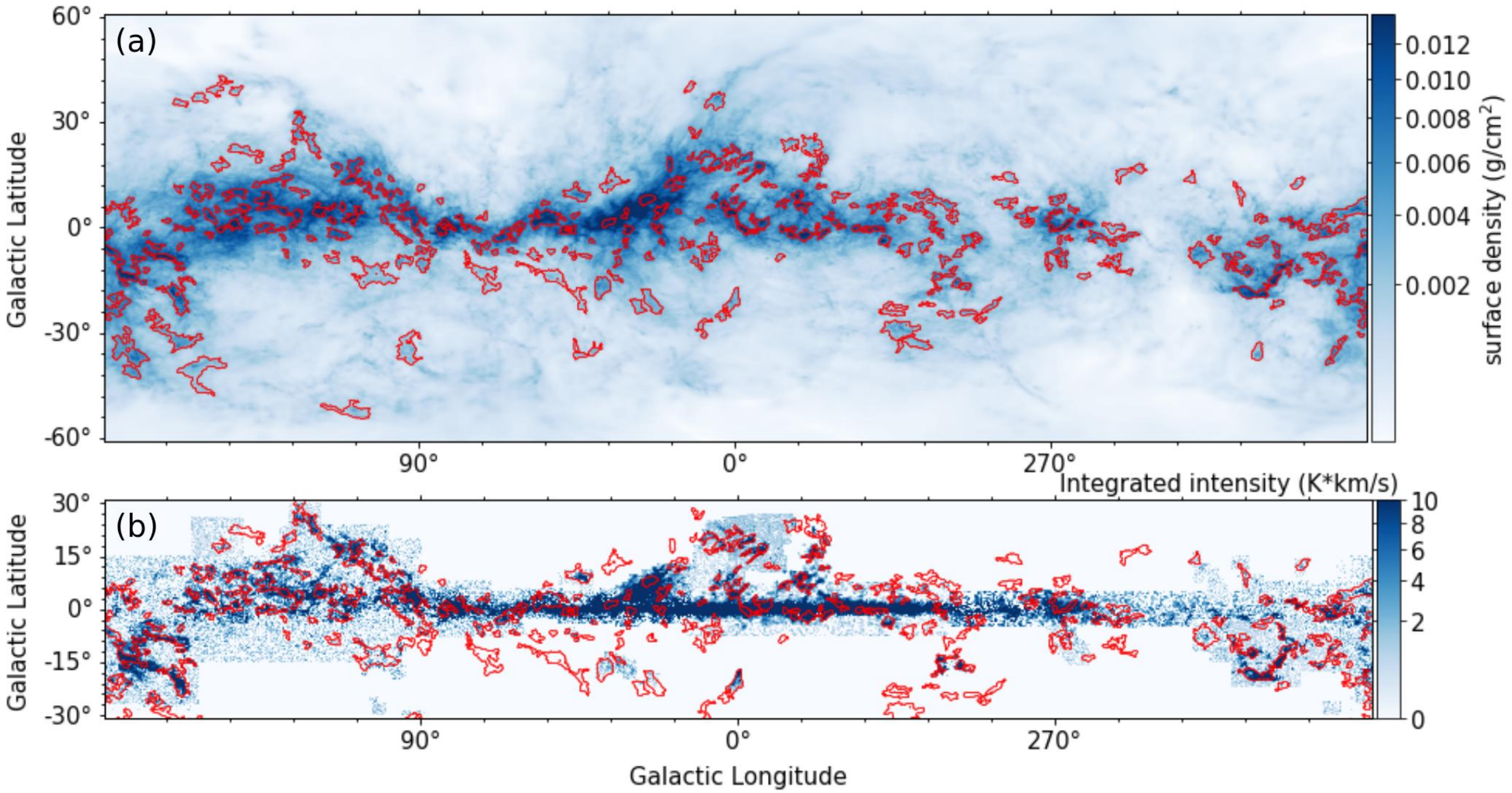}
\caption{
Structural identification.
(a) The surface density of hydrogen nuclei on the $l$-$b$ plane. Red contours are the structures identified by the dendrogram algorithm; (b) The integrated intensity map of CO (1-0) line emission. Red contours are the same with panel (a).}
\label{large}
\end{figure*}
\begin{figure}
\centering
\includegraphics[width=0.45\textwidth]{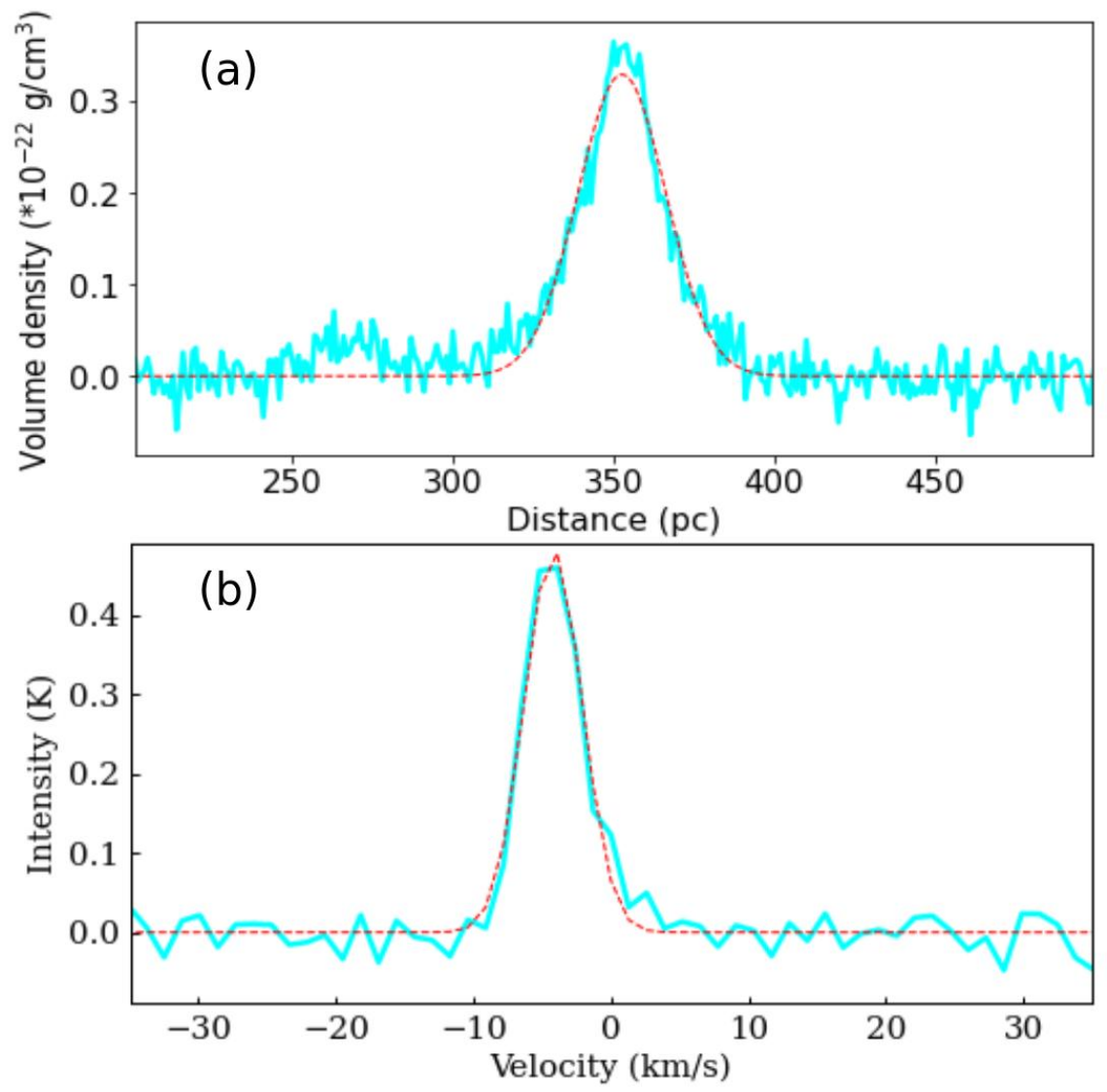}
\caption{The average "spectra" (cyan lines) of a single structure extracted from the $l$-$b$-$v$ and $l$-$b$-$d$ cubes, respectively. (a) The averaged intensity distribution along the velocity axis; (b) The averaged density distribution along the distance axis. Red dashed lines represent the Gaussian fitting.}
\label{spectra}
\end{figure}

Fig.\ref{large}(a) shows the surface density map of hydrogen nuclei on the $l$-$b$ plane. We applied the dendrogram algorithm to identify the local dense structures based on this surface density map. Then the CO (1-0) cube was utilized  to derive the velocity dispersion of the identified structures. Since the
CO (1-0) cube is in the Galactic coordinates ($l$-$b$-$v$), we also used the 3D dust map provided in the Galactic coordinate system ($l$-$b$-$d$)\footnote{\url{https://zenodo.org/records/10658339}}. This allows both cubes to be analyzed by the same way, as they have the same data structure.

As described in \citet{Rosolowsky2008-679}, the dendrogram algorithm decomposes density data into hierarchical structures. Using the {\it astrodendro} package \footnote{\url{https://dendrograms.readthedocs.io/en/stable/index.html}}, there are three major input parameters for the dendrogram algorithm: {\it min\_value} for the minimum value to be considered in the dataset, {\it min\_delta} for a leaf that can be considered as an independent entity, and {\it min\_npix} for the minimum area of a structure.
In this work, we take the values of {\it min\_value}= 5*$\Sigma_{\rm rms}$, {\it min\_delta} = 5*$\Sigma_{\rm rms}$, where $\Sigma_{\rm rms}$ is the average density of the background on the map shown in Fig.\ref{large}(a). After trying different values, we finally set {\it min\_npix} = 100 pixels to ensure that the identified structures are not too small.
As shown in Fig.\ref{large}(a), the structures identified by the dendrogram correspond well to the surface density distribution.
In Fig.\ref{large}(b), 
CO and dust also show a good correspondence in their distribution, but CO emission is primarily concentrated in the Galactic plane.

Using the mask of each identified structure in Fig.\ref{large}(a), we extracted the average "spectra" according to the $l$-$b$-$v$ and $l$-$b$-$d$ cubes. 
Specifically, for each identified structure, we averaged the intensity or density values across all pixels within the masked region and analyzed the intensity distribution along the velocity axis and the density distribution along the distance axis, as shown in Fig.\ref{spectra}. This approach allows us to investigate the characteristic velocity and distance distributions of the structures, providing insights into their kinematics and spatial properties.
If an identified structure is a single structure without overlap, we would expect both types of "spectra" to exhibit a single peak, as shown in Fig.\ref{spectra}. In total, 90 structures satisfy this condition. 
From the "spectra", we can obtain the systematic velocity, the velocity dispersion ($\sigma_{g}$), the distance and the thickness of each structure. The thickness is the full width at half maximum (FWHM) of the line profile, displayed in Fig.\ref{spectra}(a).
The dendrogram algorithm approximates the morphology of each structure as an ellipse. Using the long and short axes ($l_{1}$ and $l_{2}$) of the effective ellipse and the half-thickness ($l_{3}$), we can estimate a 3D effective radius for each structure, $R\approx(l_{1}*l_{2}*l_{3})^{1/3}$.

\subsection{Tidal strength}

In Sec.\ref{calculation}, for a subregion of the 3D map, based on the 3D density distribution, the gravitational potential was computed by solving the Poisson equation, then the tidal tensor was derived from the gravitational potential. For a cloud, its tidal tensor can be divided into two parts (internal and external) due to the matter inside and outside the cloud. Different decomposition methods result in four possible maximum eigenvalues ($|\lambda_{\rm ext, max}|$) of the external tidal tensor ($\mathrm{\mathbf{T}_{ext}}$). An independent and original pixel-by-pixel computation method was used to select the optimal $|\lambda_{\rm ext, max}|$. For the eigenvalues, we found that the optimal $\mathrm{\mathbf{T}_{ext}}$
shows the same formalism with the tidal tensor of a point-mass. Therefore, for the external tides exerted by external material at a point within a cloud, the external material can be approximated as a collection of point masses. Each point mass generates its own tidal field. The combined effect of the external tides is equivalent to the superposition of the tidal fields produced by all these point masses. Essentially, this approach is analogous to the pixel-by-pixel computation. As shown in Fig.\ref{compare},
the average tidal strength in the clouds calculated by the two methods shows comparable results.

Our current computational resources are insufficient to directly calculate the tidal tensor of the entire 3D map. Therefore, we adopted the pixel-by-pixel computation approach to calculate the average tidal strength ($\lambda_{\rm pp}$) of all structures in the subsequent analysis. This method only considers the structure being calculated and its neighboring matter each time. Typically, accounting for matter within 8–10 times the structure's radius is sufficient, as tidal strength decreases rapidly with distance (equation.\ref{point}).

\subsection{Velocity dispersion}

\begin{figure}
\centering
\includegraphics[width=0.45\textwidth]{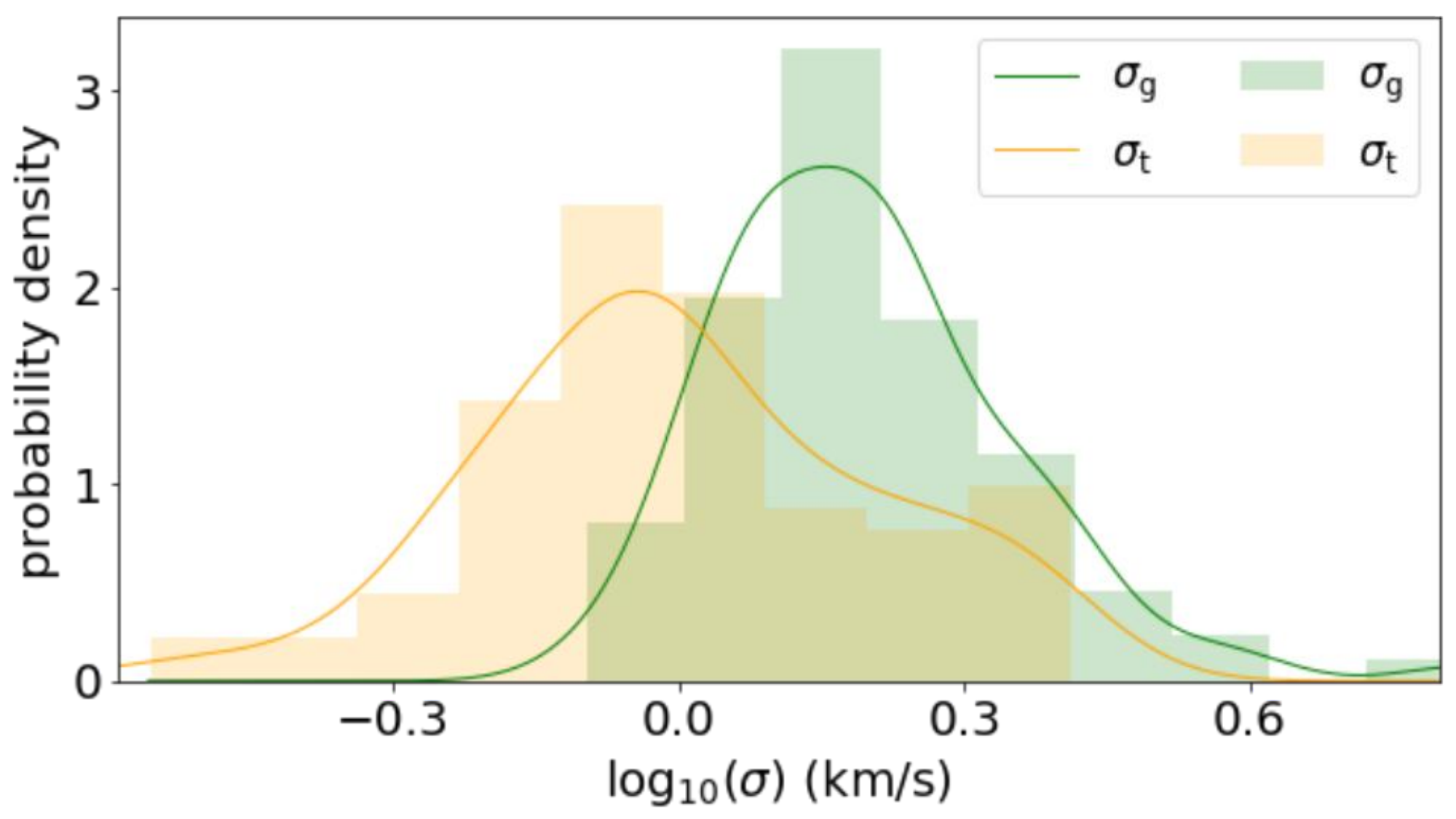}
\caption{The KDE (kernel density estimation) and the histogram of the velocity dispersion of the  clouds. $\sigma_{t}$ is the equivalent velocity dispersion derived from the average tidal strength. $\sigma_{g}$ is the velocity dispersion measured from CO emission.}
\label{dispersion}
\end{figure}

We estimated an equivalent velocity dispersion for the clouds according to the average tidal strength, as done in \citet{Zhou2024-686-146}.
The tidal strength defined in this work is the tidal acceleration per unit distance \citep{Stark1978-225}. Thus, for a cloud with radius $R$ and mass $M$, $\lambda_{\rm pp}*R$ is the tidal acceleration. 
The velocity dispersion of a gas structure measures the intensity of internal gas motion within the structure. The tides induced by external gravity pervade the interior of the gas structure. We propose that these ubiquitous gravitational differences serve as the driving force for the motion of gas parcels within the structure. 
The work done by tidal forces is converted into kinetic energy, which corresponds to the equivalent velocity dispersion, $\sigma_{t}$, i.e.
\begin{equation}
\frac{1}{2}M\sigma_{t}^{2} \approx \int_{0}^{R} M \lambda_{\rm pp} R \, dR.
\end{equation}
Then, we have
\begin{equation}
\sigma_{t} \approx \sqrt{\lambda_{\rm pp}*R^2}.
\end{equation} 
Since $\sigma_{t}$ is derived from the average tidal strength over the entire structure, it is isotropic and can be directly compared to the observed velocity dispersion of the structure obtained in Sec.\ref{identification}.
As shown in Fig.\ref{dispersion}, $\sigma_{t}$ is roughly comparable with $\sigma_{g}$ in magnitude, which means that tidal effects from external gravity can significantly contribute to the velocity dispersion of the clouds. A relatively smaller value of $\sigma_{t}$ is also reasonable, as other physical processes (such as gravitational collapse and feedback activities) also contribute to the overall velocity dispersion of the cloud.

\section{Discussion and conclusions}

The 3D density distribution derived from the 3D dust map is crucial for calculating the tidal field. However, velocity information is only available from the projected CO (1-0) line emission. To address this, we first projected the 3D density cube onto the $l$-$b$ plane and then applied the dendrogram algorithm to identify the clouds. As shown in Fig.\ref{large}, the distributions of projected CO and dust emission exhibit a good correspondence. 
For 90 non-overlapping clouds, we fitted the systematic velocity, velocity dispersion, distance, and thickness based on the average “spectra” extracted from the $l$-$b$-$v$ and $l$-$b$-$d$ cubes. These clouds were then used to study tidal effects. We employed two methods to estimate tidal strength: tidal tensor analysis and pixel-by-pixel computation, both of which produced comparable results. The equivalent velocity dispersion, derived from the average tidal strength, is roughly consistent in magnitude with the observed velocity dispersion, suggesting that tidal effects from external gravity play a significant role in shaping the velocity dispersion of the clouds.
The significant impact of external gravity from the neighboring material on the physical properties of molecular clouds has also been demonstrated in \citet{Ramirez2022-515,Li2024-528,Zhou2024-686-146,Zhou2025-537a}. The extensive tides can produce turbulence, suppresses fragmentation and slow down the gravitational collapse.
Extended gas structures in the ISM are sensitive to the tidal effects exerted by surrounding material. Future studies should incorporate these tidal effects into star formation theories. Star formation in molecular clouds typically occurs within locally dense clumps, it is essential to study these clumps in the context of their surrounding environment rather than in isolation, as their interactions with the surrounding material cannot be ignored.

As shown in Sec.\ref{calculation}, the calculation of tidal strength is based on the density.
In this work, we adopted the method described in \citet{O'Neill2024-973} to convert the differential extinction in the 3D dust map to the density of hydrogen nuclei. However, this method relies on certain assumptions, such as a constant ratio of hydrogen column density to extinction. To achieve a more accurate calculation of tidal strength, better estimates of both the density distribution and magnitude are essential.



 
\bibliographystyle{aa} 
\bibliography{ref}

\appendix

\section{Tidal calculation}\label{calculation}

\subsection{Tidal tensor}\label{tensor}

\begin{figure}
\centering
\includegraphics[width=0.5\textwidth]{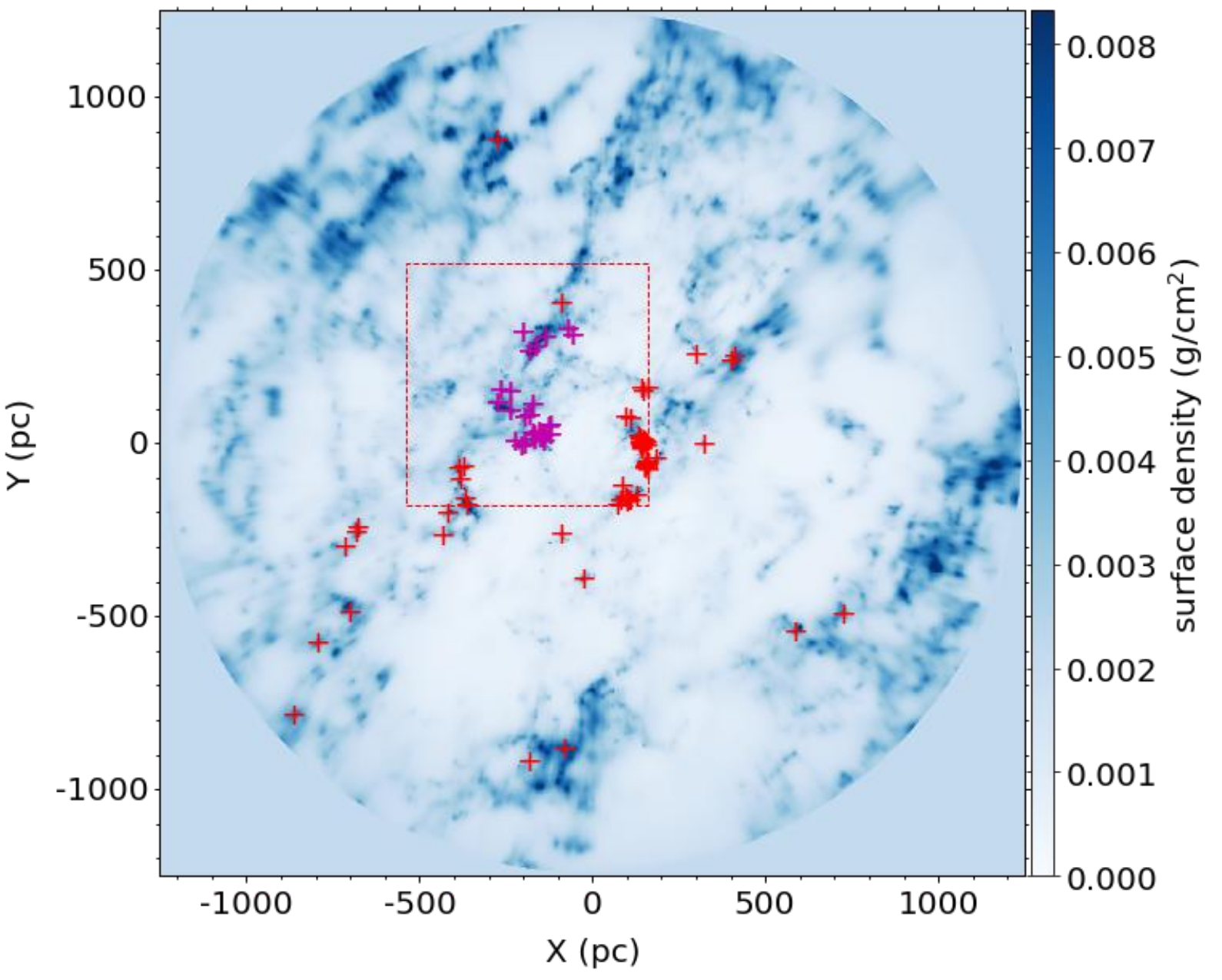}
\caption{The surface density of hydrogen nuclei on the $XY$ plane. The pluses represent the 90 non-overlapping structures in Sec.\ref{identification}.}
\label{sub}
\end{figure}

\begin{figure*}
\centering
\includegraphics[width=0.95\textwidth]{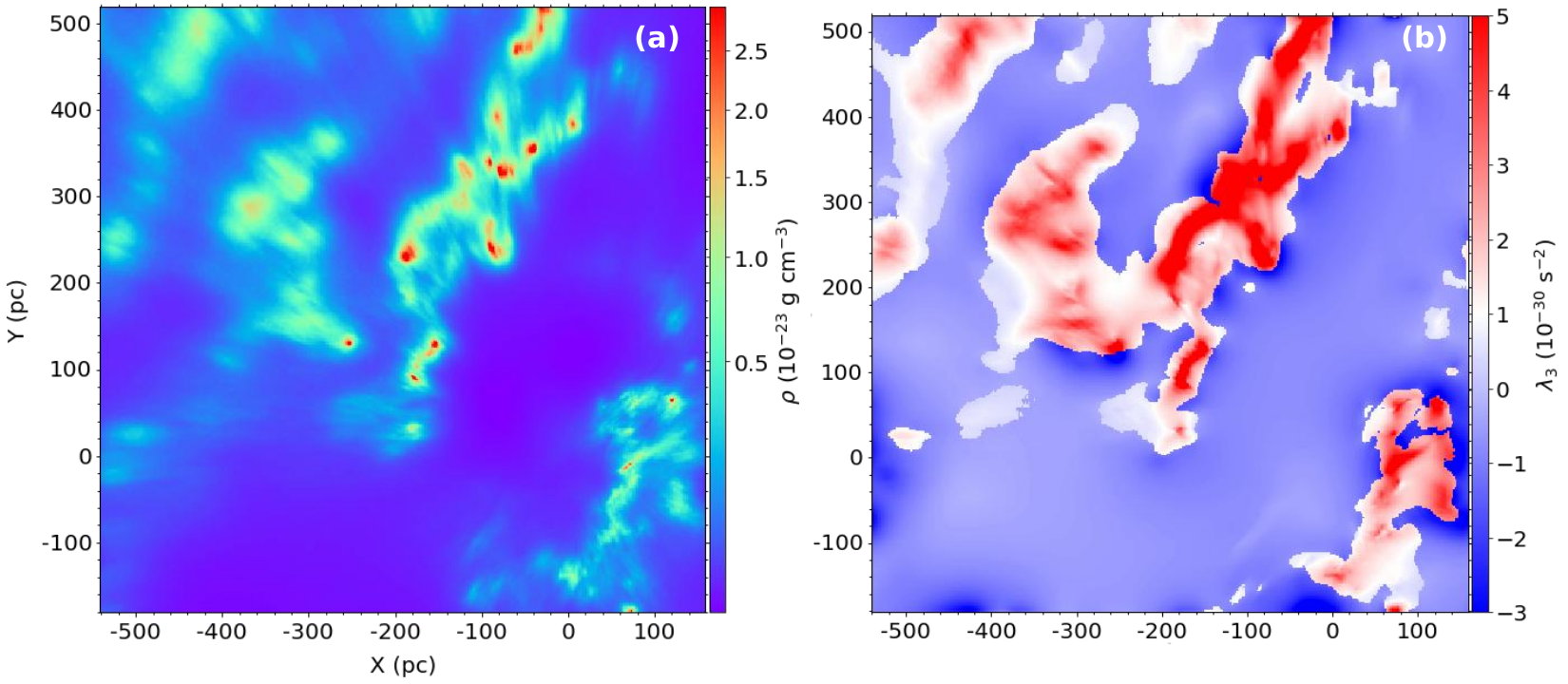}
\caption{
The tidal field in the Heliocentric Galactic Cartesian coordinate system is calculated based on the 3D dust distribution near the Sun. The Sun is located at the origin (0,0). Here, we present snapshots of the volume density and the maximum eigenvalue of the tidal tensor on the $XY$ plane at $Z=80$ pc.}
\label{snap}
\end{figure*}

In the tidal tensor computation,
we utilized the 3D dust map available in the Heliocentric Galactic Cartesian coordinate system ($X$-$Y$-$Z$)\footnote{\url{https://zenodo.org/records/10658339}}.
Based on the 3D density distribution derived from the 3D dust map, the gravitational potential is computed by solving the Poisson equation
$\nabla^2 \phi = 4 \pi G \rho$ in the Fourier space,
\begin{equation}
\Phi_{k, \rm 3D} = -\frac{4\pi G\rho_k}{k_{\rm 3D}^2}\;,
\label{3D}
\end{equation} 
where we first calculate $\rho_k = \hat{f}(\rho(x, y, z))$. Then the gravitational potential in the real space is computed through $\Phi_(x, y, z) = \hat{f}^-1(\Phi_k)$. $\hat{f}$ and $\hat{f}^{-1}$ are the Fourier transform and the inverse Fourier transform. For a gravitational potential field $\Phi$, the tidal tensor $\mathbf{T}$ is defined as
\begin{equation}
    \mathrm{T}_{ij} = \partial_i \partial_j \Phi.
\label{tij}
\end{equation}
%
The tidal tensor is symmetric and real-valued. Thus, it can be written in orthogonal form.
The ordered and diagonalized tidal tensor is given as, 
\begin{gather}
    \mathbf{T}
    =
    \begin{bmatrix}
    \lambda_1 &0&0\\
    0&\lambda_2&0\\
    0&0&\lambda_3
    \end{bmatrix}.
\end{gather}
The eigenvalues of the tidal tensor provide insight into gravitationally induced deformations, such as compression or disruption, at a given point within the gravitational field. The sign of an eigenvalue determines whether the corresponding mode is compressive ($\lambda_i > 0$) or disruptive ($\lambda_i < 0$), while its magnitude indicates the strength of the respective compressive or disruptive effect.
The "ordered" means $|\lambda_3|>|\lambda_2|>|\lambda_1|$.
The trace of the tidal tensor, $\mathrm{Tr(\mathbf{T})}=\sum_{i=1}^{3} \lambda_i$, contains the local density information in the Poisson equation:
\begin{equation}
    \mathrm{Tr(\mathbf{T})} = \nabla^2 \Phi = 4\pi G \rho.
\end{equation}
This implies that the trace is 0 if the point at which the tidal tensor is evaluated outside the mass distribution. 

Due to computational resource limitations, we focused our calculations on the subregion ($X \times Y \times Z$ = 700 pc $\times$ 700 pc $\times$ 700 pc) outlined by the red box in Fig.\ref{sub}. 
Fig.\ref{snap} displays a snapshot of the density and the maximum eigenvalue of the tidal tensor ($\lambda_3$) on $XY$ plane, when $Z=80$ pc.

As discussed in \citet{Ganguly2024-528}, 
the tidal tensor can be decomposed into three components: the tidal tensor generated solely by the matter within the structure, $\mathbf{T}_{\mathrm{int}}$; the tidal tensor arising exclusively from the external matter distribution, $\mathbf{T}_{\mathrm{ext}}$; and their sum, which represents the contribution from the entire matter distribution:
\begin{equation}
\mathbf{T}_{\mathrm{tot}}=\mathbf{T}_{\mathrm{int}}+\mathbf{T}_{\mathrm{ext}}.
\label{3t}
\end{equation}
$\mathrm{\mathbf{T}_{tot}}$ represents the net deformation due to both the structure itself and its external material. 
The trace of the three parts are:
\begin{gather}\label{eq:trace1}
    \mathrm{Tr(\mathbf{T}_{tot})} = 4\pi G \rho\\
    \mathrm{Tr(\mathbf{T}_{int})} = 4\pi G \rho\\
    \mathrm{Tr(\mathbf{T}_{ext})} = 0.
    \label{trace3}
\end{gather}
The trace in equation.\ref{trace3} suggests that $\mathrm{\mathbf{T}_{ext}}$ must include both compressive and disruptive modes, corresponding to positive and negative $\lambda_i$, respectively. In contrast, $\mathrm{\mathbf{T}_{int}}$ and $\mathrm{\mathbf{T}_{tot}}$ must contain at least one compressive mode but can also be entirely compressive.
According to equation.\ref{3t},
the 3D tidal tensor can be decomposed as
\begin{equation}
\begin{bmatrix}
    \lambda_1 &0&0\\
    0&\lambda_2&0\\
    0&0&\lambda_3
    \end{bmatrix} =  \begin{bmatrix}
    A_{0} &0&0\\
    0&A_{1}&0\\
    0&0&A_{2}
    \end{bmatrix} +  \begin{bmatrix}
    A_{3} &0&0\\
    0&A_{4}&0\\
    0&0&A_{5}
    \end{bmatrix}.
    \label{decompose}
\end{equation}
For the 6 unknown parameters, they satisfy the following four equations:
\[
\left\{
\begin{array}{l}
A_{0} + A_{3} = \lambda_1 \\
A_{1} + A_{4} = \lambda_2 \\
A_{2} + A_{5} = \lambda_3 \\
A_{3} + A_{4} + A_{5} = \lambda_1 + \lambda_2 + \lambda_3 
\end{array}
\right.
\]
Since the 6 unknown parameters are real numbers, we set $A_{4}=k*A_{5}$ and $A_{1}=h*A_{2}$, where $k$ and $h$ are arbitrary constants, then we have the solutions of the unknown parameters:
\[
\left\{
\begin{array}{l}
 A_{0} = \frac{h k \lambda_{3} - h \lambda_{2} + k \lambda_{3} - \lambda_{2}}{h - k} \\  
 A_{1} = \frac{- h k \lambda_{3} + h \lambda_{2}}{h - k} \\ 
 A_{2} = \frac{- k \lambda_{3} + \lambda_{2}}{h - k} \\ 
 A_{3} = \frac{- h k \lambda_{3} + h \lambda_{1} + h \lambda_{2} - k \lambda_{1} - k \lambda_{3} + \lambda_{2}}{h - k} \\
 A_{4} = \frac{h k \lambda_{3} - k \lambda_{2}}{h - k} \\
 A_{5} = \frac{h \lambda_{3} - \lambda_{2}}{h - k} 
\end{array}
\right.
\]
The solutions require $h \neq k$. Below we consider several limiting cases:

1. When $k \gg 1$ and $h \approx 1$, the solutions of 6 unknown parameters are :
$A_{0} =- 2 \lambda_{3}$,
$A_{1} =\lambda_{3}$,
$A_{2} =\lambda_{3}$,
$A_{3} =\lambda_{1} + 2 \lambda_{3}$,
$A_{4} =\lambda_{2} - \lambda_{3}$,
$A_{5} =0$.

2. When $k \gg 1$ and $h \ll 1$, the solutions of 6 unknown parameters are :
$A_{0} =- \lambda_{3}$,
$A_{1} =0$,
$A_{2} =\lambda_{3}$,
$A_{3} =\lambda_{1} + \lambda_{3}$,
$A_{4} =\lambda_{2}$,
$A_{5} =0$.

3. When $k \approx 1$ and $h \gg 1$, the solutions of 6 unknown parameters are :
$A_{0} =- \lambda_{2} + \lambda_{3}$,
$A_{1} =\lambda_{2} - \lambda_{3}$,
$A_{2} =0$,
$A_{3} =\lambda_{1} + \lambda_{2} - \lambda_{3}$,
$A_{4} =\lambda_{3}$,
$A_{5} =\lambda_{3}$.

4. When $k \approx 1$ and $h \ll 1$, the solutions of 6 unknown parameters are :
$A_{0} =\lambda_{2} - \lambda_{3}$,
$A_{1} =0$,
$A_{2} =- \lambda_{2} + \lambda_{3}$,
$A_{3} =\lambda_{1} - \lambda_{2} + \lambda_{3}$,
$A_{4} =\lambda_{2}$,
$A_{5} =\lambda_{2}$.

5. When $k \ll 1$ and $h \gg 1$, the solutions of 6 unknown parameters are :
$A_{0} =- \lambda_{2}$,
$A_{1} =\lambda_{2}$,
$A_{2} =0$,
$A_{3} =\lambda_{1} + \lambda_{2}$,
$A_{4} =0$,
$A_{5} =\lambda_{3}$.

6. When $k \ll 1$ and $h \approx 1$, the solutions of 6 unknown parameters are :
$A_{0} =- 2 \lambda_{2}$,
$A_{1} =\lambda_{2}$,
$A_{2} =\lambda_{2}$,
$A_{3} =\lambda_{1} + 2 \lambda_{2}$,
$A_{4} =0$,
$A_{5} =- \lambda_{2} + \lambda_{3}$.

Assuming that the
largest eigenvalue of $\mathrm{\mathbf{T}_{ext}}$ dominates the deformation of a given structure due to the external tides, in the 6 limiting cases, the
maximum eigenvalue of $\mathrm{\mathbf{T}_{ext}}$ (i.e. $\lambda_{\rm ext, max}$) can be: $|- 2 \lambda_{2}|$, $|- 2 \lambda_{3}|$, $|\lambda_{2} - \lambda_{3}|$, $|- \lambda_{2}|$ and $|- \lambda_{3}|$. Since the eigenvalues can become negative, we consider their absolute values in order to compare the relative magnitude. Similar to the case of $A_{4}=k*A_{5}$ and $A_{1}=h*A_{2}$, there are other 8 situations: 

$A_{0}=h*A_{1}$ and $A_{3}=k*A_{4}$, 

$A_{0}=h*A_{1}$ and $A_{3}=k*A_{5}$, 

$A_{0}=h*A_{1}$ and $A_{4}=k*A_{5}$, 

$A_{0}=h*A_{2}$ and $A_{3}=k*A_{4}$,

$A_{0}=h*A_{2}$ and $A_{3}=k*A_{5}$, 

$A_{0}=h*A_{2}$ and $A_{4}=k*A_{5}$, 

$A_{1}=h*A_{2}$ and $A_{3}=k*A_{4}$, 

$A_{1}=h*A_{2}$ and $A_{3}=k*A_{5}$.

For all limiting cases, there are 12 possible $|\lambda_{\rm ext, max}|$, i.e. 
$|- 2 \lambda_{1}|$, 
$|- \lambda_{1}|$,
$|- 2 \lambda_{2}|$, 
$|- \lambda_{2}|$, 
$|- 2 \lambda_{3}|$,
$|- \lambda_{3}|$,
$|\lambda_{1} - \lambda_{2}|$,
$|\lambda_{1} - \lambda_{3}|$, 
$|\lambda_{2} - \lambda_{3}|$, 
$|\lambda_{1} - \lambda_{2}|/2$,
$|\lambda_{1} - \lambda_{3}|/2$ and
$|\lambda_{2} - \lambda_{3}|/2$.
Since $|\lambda_3|>|\lambda_2|>|\lambda_1|$, we only need to consider four of them, i.e.
$|- 2 \lambda_{3}|$, $|\lambda_{1} - \lambda_{2}|$,
$|\lambda_{1} - \lambda_{3}|$ and 
$|\lambda_{2} - \lambda_{3}|$.

\subsection{Pixel-by-pixel computation}\label{pp}
\begin{figure}
\centering
\includegraphics[width=0.35\textwidth]{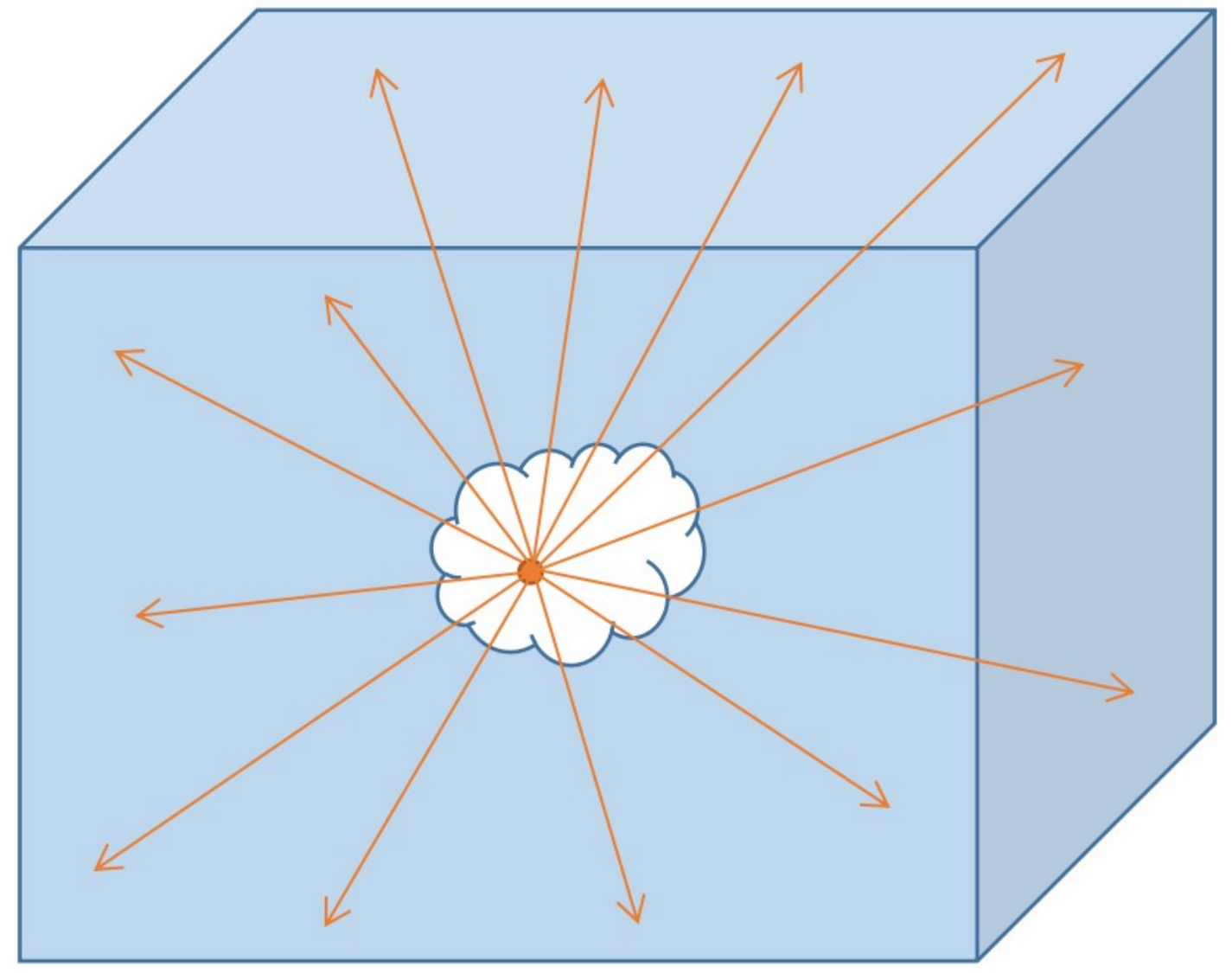}
\caption{
The pixel-by-pixel computation.
Taking out a cloud from the 3D density cube and calculating the external tides from all external material at a point in the cloud.}
\label{carton}
\end{figure}

\begin{figure}
\centering
\includegraphics[width=0.45\textwidth]{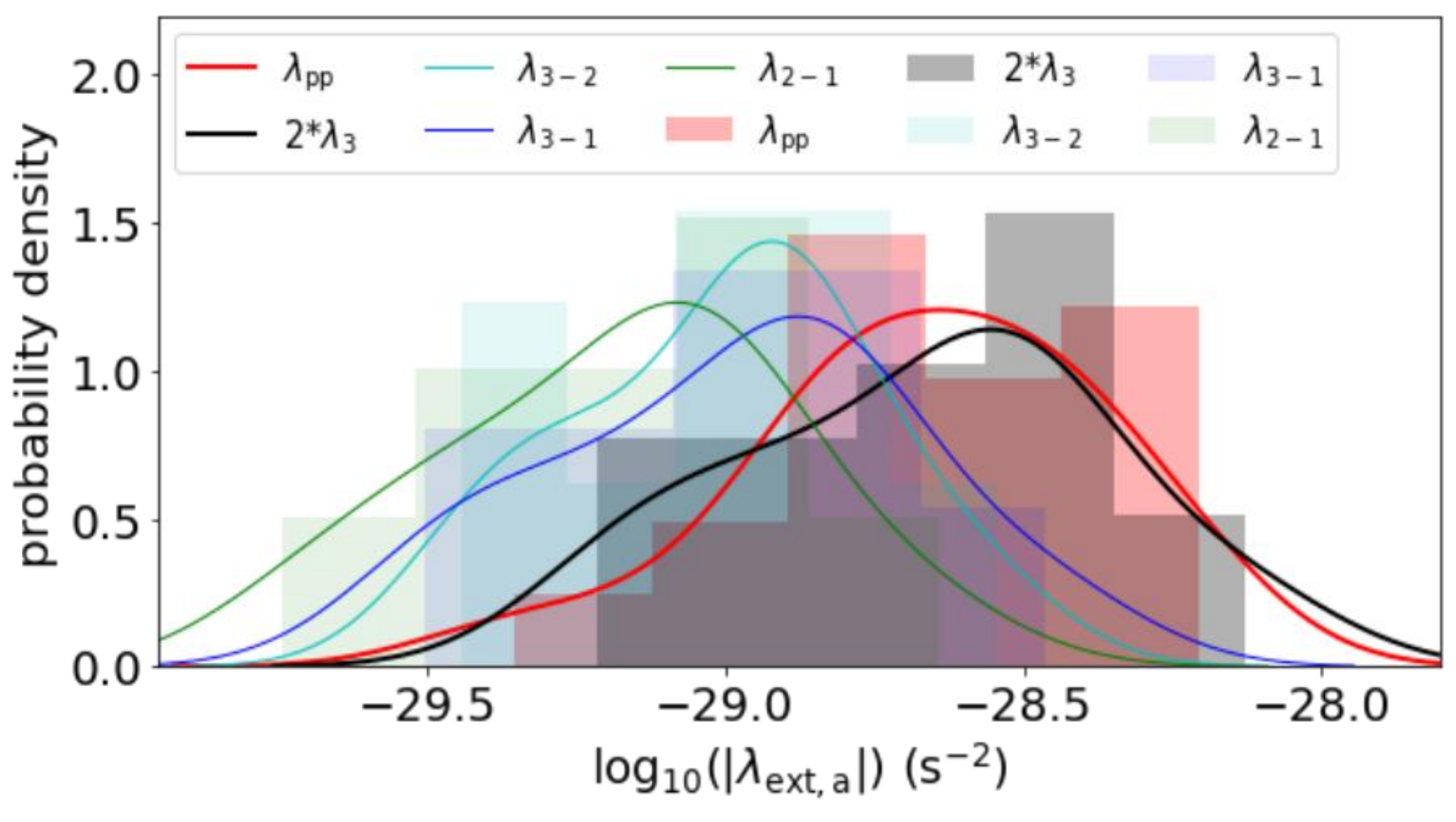}
\caption{The KDE (kernel density estimation) and the histogram of the average external tidal strength of the clouds. 
$|- 2 \lambda_{3}|$, $|\lambda_{1} - \lambda_{2}|$,
$|\lambda_{1} - \lambda_{3}|$ and 
$|\lambda_{2} - \lambda_{3}|$ are four possible maximum eigenvalues derived in Sec.\ref{tensor} and $\lambda_{\rm pp}$ is the tidal strength from the pixel-by-pixel computation.
The comparison here is only conducted on the clouds marked by magenta "+" within the box of Fig.\ref{sub}.}
\label{compare}
\end{figure}

Below, we introduce an independent and original method to select the optimal $|\lambda_{\rm ext, max}|$ from four possible ones. 
At a distance $R^{'}$ from a pixel B with mass $M^{'}$, the tidal strength sustained by a pixel A due to the external gravity of pixel B is
\begin{equation}
T \approx \frac{2GM^{'}}{R^{'3}}.
\label{point}
\end{equation}
The cumulative tidal strength at a pixel signifies the aggregate deformation resulting from external gravity at that point. Unlike a rigid body or a point-mass, a gas structure is flexible and can be deformed. Gas structures are often irregular in shape, and their morphologies can be quite intricate. While the gravitational forces acting on a gas structure from different directions might balance out, the tidal effects within the structure, driven by external gravity, do not. To quantify the collective tidal strength within a gas structure induced by external gravity, we adopt the scalar superposition.
According to the central coordinates ($X$, $Y$ and $Z$) and radius of each cloud identified in Sec.\ref{identification}, we can divide the 3D density cube into two parts, i.e. the cloud itself and all external material apart from this cloud. For the external tides from external material sustained by a point in the cloud, a simple but heavy calculation is using equation.\ref{point} to directly calculate the tidal strength at that point pixel-by-pixel through the entire cube, as illustrated in Fig.\ref{carton}. 
Fig.\ref{compare} shows the distribution of the average tidal strengths (four possible $|\lambda_{\rm ext, max}|$ and the pixel-by-pixel computed one, i.e. $\lambda_{\rm pp}$) for the  clouds marked by magenta "+" within the box of Fig.\ref{sub}. Since the clouds at the boundary cannot account for all neighboring material, they were excluded. The distribution of $|- 2 \lambda_{3}|$ is closest to $\lambda_{\rm pp}$. In equation.\ref{decompose}, when $A_{0}=- 2 \lambda_{3}$, we obtain a decomposition of the tidal tensor,
\begin{equation}
\mathbf{T}_{\mathrm{tot}}=\mathbf{T}_{\mathrm{ext}}+\mathbf{T}_{\mathrm{int}} =  \begin{bmatrix}
    -2\lambda_{3} &0&0\\
    0&\lambda_{3}&0\\
    0&0&\lambda_{3}
    \end{bmatrix} +  \begin{bmatrix}
    \lambda_{1}+2\lambda_{3} &0&0\\
    0&\lambda_{2}-\lambda_{3}&0\\
    0&0&0
    \end{bmatrix}.
    \label{decompose-f}
\end{equation}
For a point-mass with mass $M_{0}$, the tidal tensor at distance $d$ is given as:
\begin{gather}
    \mathbf{T}
    =
    \begin{bmatrix}
    -\frac{2GM_{0}}{d^3} &0&0\\
    0&\frac{GM_{0}}{d^3}&0\\
    0&0&\frac{GM_{0}}{d^3}.
    \end{bmatrix}
\end{gather}
As shown in Fig.\ref{compare}, 
$<|2\lambda_{3}|> \approx <2GM^{'}/R^{'3}>$ 
(equation.\ref{point}), and the eigenvalues of $\mathrm{\mathbf{T}_{ext}}$ in equation.\ref{decompose-f} has the same formalism with the tidal tensor of a point-mass. Therefore, 
for the external tides from external material sustained by a point in a cloud, the external material can be divided into approximate point-masses. Each point-mass can produce its own tidal field. The external tides from all external material are equivalent to the superposition of the tidal fields of all divided point-masses. In essence, it is equivalent to the above pixel-by-pixel computation. 

\begin{acknowledgements}
Thanks to the referee for the constructive comments, which have significantly improved this work. Thanks to G. X. Li for the helpful comments and discussions.
\end{acknowledgements}

\end{document}